\documentclass[conference]{IEEEtran}
\IEEEoverridecommandlockouts
\usepackage{cite}
\usepackage{amsmath,amssymb,amsfonts}
\usepackage{algorithmic}
\usepackage{graphicx}
\usepackage{textcomp}
\usepackage{xcolor}
\usepackage{soul}
\usepackage{booktabs}
\usepackage{pifont}
\usepackage{tikz}
\usepackage{balance}
\usepackage{hyperref}
\usetikzlibrary{shapes.geometric, arrows, positioning}
\tikzstyle{block} = [rectangle, rounded corners, draw=black, align=center, text width=4cm, minimum height=1.5cm]
\tikzstyle{blockcolor} = [block, fill=blue!10]
\tikzstyle{arrow} = [thick,->,>=stealth]

\def\BibTeX{{\rm B\kern-.05em{\sc i\kern-.025em b}\kern-.08em
    T\kern-.1667em\lower.7ex\hbox{E}\kern-.125emX}}

\newcommand{\minisection}[1]{\vspace{0.05in}\noindent {\bf #1}}

\begin{document}


\title{EdgeProfiler: A Fast Profiling Framework for Lightweight LLMs on Edge Using Analytical Model}

\author{%
  \IEEEauthorblockN{%
    Alyssa Pinnock\textsuperscript{1}$^*$\thanks{$^*$These authors contributed equally to this work.},
    Shakya Jayakody\textsuperscript{1}$^*$,
    Kawsher A Roxy\textsuperscript{2},
    Md Rubel Ahmed\textsuperscript{3}
  }
  \IEEEauthorblockA{\textsuperscript{1}University of Central Florida, \textsuperscript{2}Intel Corporation, \textsuperscript{3}Louisiana Tech University}
  \IEEEauthorblockA{Email: \{al310186, shakya\}@ucf.edu\textsuperscript{1}, kawsher.roxy@intel.com\textsuperscript{2}, mahmed@latech.edu\textsuperscript{3}}
}

\maketitle

\begin{abstract}

This paper introduces EdgeProfiler, a fast profiling framework designed for evaluating lightweight Large Language Models (LLMs) on edge systems. While LLMs offer remarkable capabilities in natural language understanding and generation, their high computational, memory, and power requirements often confine them to cloud environments. EdgeProfiler addresses these challenges by providing a systematic methodology for assessing LLM performance in resource-constrained edge settings.
The framework profiles compact LLMs, including TinyLLaMA, Gemma3.1B, Llama3.2-1B, and DeepSeek-r1-1.5B, using aggressive quantization techniques and strict memory constraints. Analytical modeling is used to estimate latency, FLOPs, and energy consumption. The profiling reveals that 4-bit quantization reduces model memory usage by approximately 60--70\%, while maintaining accuracy within 2--5\% of full-precision baselines. Inference speeds are observed to improve by 2--3$\times$ compared to FP16 baselines across various edge devices. Power modeling estimates a 35--50\% reduction in energy consumption for INT4 configurations, enabling practical deployment on hardware such as Raspberry Pi 4/5 and Jetson Orin Nano Super.
Our findings emphasize the importance of efficient profiling tailored to lightweight LLMs in edge environments, balancing accuracy, energy efficiency, and computational feasibility.

\end{abstract}


\section{Introduction}
\label{sec:intro}


Advances in Artificial Intelligence (AI) and deep learning have fueled remarkable progress in natural language processing (NLP), enabling applications such as summarization, text generation, question answering, and more. At the forefront of these developments are LLMs, which demonstrate an unprecedented ability to interpret and generate human-like text. LLMs have contributed to breakthroughs in mobile applications, healthcare, and situational analysis, among other domains. However, these models typically require substantial computational resources, memory, and power, which confine their deployment to cloud environments equipped with high-performance servers and GPUs.

Despite these challenges, there is growing interest in implementing LLMs on edge devices, including microcontrollers and low-power platforms. Edge deployment offers critical advantages in environments where cloud connectivity is unreliable or unavailable. For instance, LLMs can improve \textit{situational awareness} during emergencies and disaster response, where cloud infrastructure may be compromised~\cite{crisisllm}. In these scenarios, devices running LLMs can provide real-time crisis management and communication support. Additionally, on-device processing enhances data security by minimizing exposure to vulnerabilities associated with transmitting data over the internet. 

Reducing dependency on cloud connectivity is also a significant advantage. Moving LLM inference to the edge allows applications to function independently of network conditions, ensuring availability even in areas with limited or no internet access. 
For example, a local LLM could manage user notes or provide contextual assistance directly on a smartphone, regardless of network strength. Moreover, local inference significantly reduces latency, which is crucial for real-time applications. Unlike cloud-based models whose response times are affected by network stability and speed, on-device inference ensures prompt interaction, enhancing the user experience. 
However, deploying LLMs on edge devices remains challenging due to their complex architectures and high memory and power requirements~\cite{s23031279}. Models such as GPT-3 and LLaMA, with billions of parameters, are not feasible for resource-constrained environments~\cite{dettmers2022gpt3,inan2023llama}. This has led to an active area of research focused on quantization techniques, model pruning, and efficient inference strategies that adapt LLMs to the limitations of edge hardware~\cite{yee2024device}. These efforts aim to strike a balance between computational feasibility, energy efficiency, and model accuracy.

In this paper, we introduce EdgeProfiler, a fast profiling framework designed to systematically evaluate the performance of lightweight LLMs in resource-constrained environments. It shows that aggressive quantization makes lightweight LLMs practical for edge deployment by reducing memory, computation, and energy costs with only minor accuracy trade-offs.
The main contributions of this paper are as follows:
\begin{itemize}
    \item We introduce EdgeProfiler, a fast profiling framework for lightweight LLMs on edge devices.
    \item Systematic evaluation of quantization and low-bit implementations to address memory and power constraints.
    \item Analysis of strategies to enable LLM inference on microcontrollers and low-power edge platforms.
    \item Comprehensive review of recent research and experimental results on deploying LLMs in hardware-constrained environments.
\end{itemize}


\section{Background}
\label{sec:background}

\subsection{Quantization techniques for efficient LLM inference}
\noindent
Quantization refers to the process of reducing the numerical precision of model parameters and activations from high-precision FP16 / FP32 format to lower-precision representations of 8-bit integer or 4-bit integer\cite{gholami2022survey}. 

\minisection{Symmetric vs. Asymmetric.} Quantization technique can be broadly classified into \textit{symmetric} and \textit{asymmetric} schemes, depending on how they map floating-point values to lower-precision integer representations.

\textit{Symmetric Quantization}:
In symmetric quantization, zero in floating-point space is exactly representable by zero in integer space. The mapping function can be defined as:
\begin{equation}
    x_{\text{int}} = \text{round}\left(\frac{x_{\text{float}}}{s}\right)
\end{equation}
where $x_{\text{float}}$ is the original floating-point value, $x_{\text{int}}$ is the quantized integer value, and $s$ is a common positive scaling factor.
Dequantization process scales back as:
\begin{equation}
    x_{\text{float}} \approx s \times x_{\text{int}}
\end{equation}

\textit{Asymmetric Quantization}:
Asymmetric quantization adds a nonzero offset to handle data distributions not centered at zero, making the quantization function:
\begin{equation}
\label{Asy_quant}
    x_{\text{int}} = \text{round}\left(\frac{x_{\text{float}} - z}{s}\right)
\end{equation}

\noindent and the corresponding dequantization is:
\begin{equation}
    x_{\text{float}} \approx s \times x_{\text{int}} + z
\end{equation}
where $z$ is the zero-point, chosen such that zero in the integer domain corresponds to a nonzero floating-point value.

Asymmetric quantization is generally used for activations, where the dynamic range shifts during inference. Symmetric quantization is highly hardware-efficient because it eliminates the need for offset additions during inference. It is particularly well-suited for weight matrices, where the value distributions are typically centered around zero. An advantage of symmetric quantization is that it requires less memory and simpler hardware logic\cite{jacob2018quantization, nagel2106white}. However, a key disadvantage is that it can result in a higher mean squared error (MSE) compared to asymmetric quantization.


\minisection{Per-Tensor vs. Per-Channel Quantization.}
Quantization can be applied either globally across an entire tensor or separately across individual channels. These two strategies, per-tensor quantization and per-channel quantization, trade off between simplicity and representational accuracy\cite{nagel2106white}.

\textit{Per-Tensor Quantization}: In per-tensor quantization, a single scale factor and, optionally, a single zero-point are used to quantize the entire tensor. The same quantization parameters are applied uniformly across all elements. similar to the equation \ref{Asy_quant}. Per-tensor quantization offers the advantages of simple implementation and high efficiency, particularly for hardware accelerators that prefer uniform scaling across all elements. However, it also has notable disadvantages: it fails to capture the variance across different channels or feature maps, and it can lead to larger quantization errors if some channels have significantly wider or narrower value ranges compared to others\cite{banner2019post}.

\textit{Per-Channel Quantization}: In per-channel quantization, each channel of each output feature in a convolutional layer or each row in a matrix has its own scale and zero-point. This allows finer control over the quantization process.
\begin{equation}
    x_{\text{int}, c} = \text{round}\left( \frac{x_{\text{float}, c} - z_c}{s_c} \right)
\end{equation}
where \( s_c \) and \( z_c \) are the scale and zero-point specific to channel \( c \).
Per-tensor quantization is preferable when memory bandwidth and hardware simplicity are prioritized. In contrast, per-channel quantization provides better model accuracy, particularly for deep neural networks where channel-wise variation is significant. Modern LLM quantization schemes often apply per-channel quantization to weights and per-tensor quantization to activations, balancing accuracy and inference efficiency\cite{yao2022zeroquant}.

\begin{figure}[t]
\centerline{\includegraphics[width=.48\textwidth]{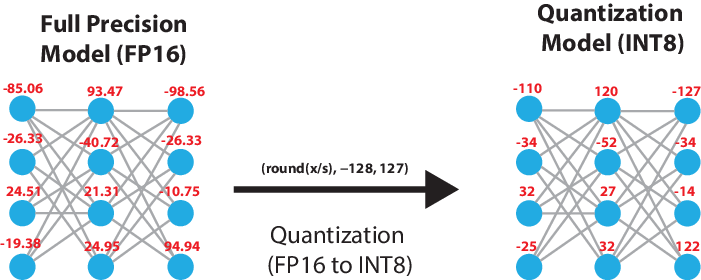}}
\caption{Quantization process from FP16 to INT8, demonstrating how a weight like 93.47 is scaled and rounded to 120 in INT8. This reduces memory size and computation while introducing minimal error, as the scaling factor preserves relative weight magnitudes.}
\vspace{-10pt}

\label{gpt3_fig}
\end{figure}

\minisection{\underline{Q}uantization-\underline{A}ware \underline{T}raining (QAT).}
Quantization-Aware Training is a technique where quantization effects are simulated during the training phase itself, allowing the model to adapt to low-precision representations. During QAT, both the forward and backward passes emulate quantized operations, typically using fake quantization functions that maintain gradient flow\cite{jacob2018quantization}. This enables the model to learn parameters that are more robust to quantization noise, resulting in significantly better accuracy compared to naive post-training quantization\cite{esser2019learned}. During QAT, the model optimizes following objective:
\begin{equation}
\min_{\theta} \; \mathbb{E}_{(x, y) \sim \mathcal{D}} \left[ \mathcal{L} \left( Q(f(x; \theta)), y \right) \right]
\end{equation}
where \( f(x; \theta) \) denotes the model with parameters \( \theta \), \( Q(\cdot) \) denotes the quantization function, \( \mathcal{L}(\cdot, \cdot) \) is the loss function, and \( \mathcal{D} \) is the training data distribution.

QAT often uses lower bit-widths 8-bit or 4-bit during simulation while keeping high-precision master weights for gradient updates. By introducing quantization noise during training, QAT yields models that maintain higher accuracy after deployment, especially important for aggressive compression regimes in edge and embedded systems\cite{dettmers2022gpt3}. Despite promising benefits, LLM quantization presents unique challenges, such as attention layers and softmax output are particularly sensitive to quantization noise. Even if inputs are quantized to 4-bit/8-bit, intermediate results often require higher precision to avoid numerical instability. Some input sequences containing rare or domain-specific tokens can significantly stress low-precision models\cite{yao2022zeroquant,frantar2022gptq, qin2024empirical}. Fig.~\ref{gpt3_fig} shows the Symmetric quantization technique.


\begin{table*}[t]
\centering
\caption{Comparison of specifications for Raspberry Pi 4, Raspberry Pi 5, and Jetson Orin Nano Super devices.}
\label{tab:device-comparison-os}
\resizebox{!}{!}{
\begin{tabular}{lcccc}
\toprule
\textbf{Device} & \textbf{CPU} & \textbf{RAM} & \textbf{Storage} & \textbf{OS} \\
\midrule
Raspberry Pi 4       & Quad-core Cortex-A72 @1.5 GHz          & 2 GB/4 GB/8 GB LPDDR4        & microSD card                                & Linux              \\
Raspberry Pi 5       & Quad-core Cortex-A76 @2.4 GHz          & 4 GB/8 GB/16 GB LPDDR4X      & microSD card, PCIe storage                  & Linux              \\
Jetson Orin Nano Super & 6-core Cortex-A78AE @1.7 GHz           & 8 GB 128-bit LPDDR5          & microSD card, PCIe storage     & Linux \\
\bottomrule
\end{tabular}
}
\vspace{-10pt}
\end{table*}

\begin{figure}[t]
\hspace{5pt}
\centerline{\includegraphics[width=0.45\textwidth]{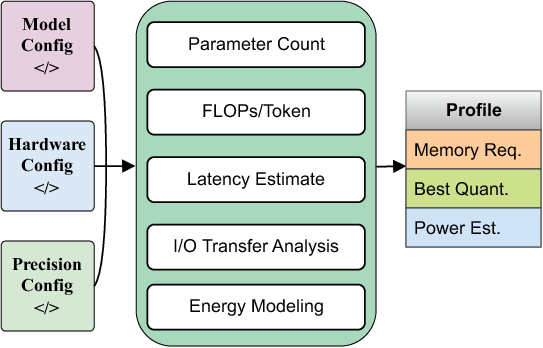}}
\caption{High-level overview of the EdgeProfiler. The framework integrates model, hardware, and precision parameters to estimate latency, memory footprint, and energy consumption, enabling exploration of performance trade-offs on edge platforms.}
\vspace{-10pt}
\label{fig:overview}
\end{figure}

\section{Related Works}
\label{sec:literature}
\minisection{Lightweight LLMs.} MobileLLM~\cite{mobilellm} demonstrates that sub-billion parameter models can provide effective NLP functionality on mobile and edge platforms, achieving a balance between resource constraints and utility. BioMistral-7B~\cite{biomistral} applies quantization and model merging to tailor LLMs for biomedical tasks, combining compression with domain-specific optimization for constrained environments.
Memory innovations, exemplified by ``LLM in a Flash”~\cite{flashllm}, reduce DRAM usage by dynamically loading only essential model weights from flash storage during inference. This approach addresses limitations of static weight loading, improving LLM feasibility for edge deployment. Another recent advancement in efficient LLM deployment is the work on automatic INT4 weight-only quantization and optimized CPU inference, as described by the authors in~\cite{shen2023efficient}.

\minisection{LLM Inference on Edge Devices.} Recent research has focused on adapting LLMs for edge devices, addressing challenges like limited memory, compute capacity, and energy constraints. Work~\cite{empirical2025} empirically studies how resource-constrained computing environments affect the design choices for personalized LLMs, uncovering key trade-offs and offering practical guidelines for customization and deployment on edge devices. However, the empirical study might be limited to specific models and hardware, affecting generalizability. Deeploy~\cite{deeploy} introduces a compiler framework that translates pre-trained Foundation Models into compact Small Language Models (SLMs) for microcontrollers, showcasing a practical approach for edge NLP tasks. 
Post-training quantization techniques, such as GPTQ ~\cite{frantar2022gptq}, compress model weights to as low as 2--4 bits, balancing memory efficiency and accuracy. Complementary methods like AWQ~\cite{lin2024awq} and SmoothQuant~\cite{xiao2023smoothquant} further enhance memory and performance trade-offs. 


\section{Methodology}
\label{sec:method}


\subsection{Performance Modeling Framework}
\noindent
At the core of our framework is the EdgeProfile, which takes three inputs: model configuration, hardware configuration, and precision configuration, as shown in the figure. \ref{fig:overview}.

\minisection{Model configuration.} Specifies transformer architecture parameters: number of layers ($L$), hidden dimension ($H$), intermediate dimension ($I$), attention heads, vocabulary size ($V$), and sequence length ($S$).

\minisection{Hardware configuration.} Encapsulates peak computational throughput (FLOPs/sec), DRAM bandwidth, storage throughput, PCIe host‐to‐device bandwidth, interconnect/network bandwidth, utilization factors ($U_\text{compute}$, $U_\text{memory}$, $U_\text{storage}$, $U_\text{H2D}$, $U_\text{net}$), and energy cost per flop or per byte accessed.

\minisection{Precision configuration.} Defines the data‐type size $B$ in bytes (4B for FP32, 2B for FP16, 1B for INT8).

\noindent
\minisection{EdgeProfiler computes}:
\paragraph{Parameter count}
The total number of model parameters is fundamental for determining storage, initialization time, and memory footprint. EdgeProfiler calculates it by summing contributions from all projection and embedding matrices across the transformer architecture:
\begin{equation}
\label{weight_embeddings}
  P = L\,(4H^2) + L\,(2HI) + 2VH
\end{equation}

Equation \ref{weight_embeddings} computes the total number of model parameters by summing contributions from self-attention, feed-forward, and embedding layers. The first term, $L(4H^2)$, counts four linear projection matrices (query, key, value, output) per layer, each of size $H \times H$. The second term, $L(2HI)$, counts the two linear layers in each feed-forward network with input dimension $H$ and expansion dimension $I$. The final term, $2VH$, accounts for input and output embedding matrices that project vocabulary tokens to hidden representations and back. This formulation allows EdgeProfiler to accurately capture total model size, critical for estimating storage requirements, weight loading times, and parameter transfer bandwidth needs during initialization and deployment.

\paragraph{FLOPs per token}
To estimate the compute workload per token, EdgeProfiler calculates the total floating-point operations required for processing:
\begin{equation}
\label{flop}
  \begin{aligned}
    \mathrm{FLOPs/token}
    &= L\bigl(6H^2 + 4HS + 4HI + 4IH + 9H\bigr)
  \end{aligned}
\end{equation}

Equation \ref{flop} calculates the total floating-point operations required to process a single token. The $6H^2$ term aggregates four projections for attention and two residual projections, while $4HS$ models token-wise dot-product attention computation over sequence length $S$. The terms $4HI + 4IH$ capture feed-forward block matmuls, and $9H$ accounts for LayerNorm, bias additions, and other elementwise operations. This equation enables precise estimation of compute workload per token, which is essential for understanding whether inference is compute-bound or memory-bound on a target edge device.

\paragraph{Memory footprint}
Estimating total memory usage is essential for ensuring inference fits within device capacity:
\begin{equation}
\label{memory}
  M = P\cdot B + S\cdot H\cdot B + 2L\cdot S\cdot H\cdot B
\end{equation}

Equation \ref{memory} estimates total memory usage, combining model weights ($P \cdot B$), hidden-state activations ($S \cdot H \cdot B$), and cached key/value tensors ($2L \cdot S \cdot H \cdot B$). Here, $B$ is bytes per parameter. The key/value cache term reflects storage of projected keys and values for each token in each layer, enabling efficient autoregressive decoding. 
This memory estimation is vital for determining whether an edge GPU or accelerator has sufficient on-chip SRAM or VRAM to store the model without offloading to external DRAM, which incurs additional latency and energy. Accurate memory modeling guides system designers in choosing appropriate batch sizes, quantization schemes, and layer fusion strategies to fit within device constraints.

\subsection{Latency Breakdown}

\paragraph{Compute latency}
EdgeProfiler estimates compute-bound inference time as:
\begin{equation}
\label{comp}
  T_{\mathrm{comp}} =
  \frac{\mathrm{FLOPs/token}}{\mathrm{peak}_{\mathrm{flops}} \times U_{\mathrm{compute}}}
\end{equation}

Equation \ref{comp} estimates inference time under compute-bound conditions by dividing total FLOPs per token by the effective device compute throughput. $\mathrm{peak}_{\mathrm{flops}}$ is the advertised peak performance of the accelerator, and $U_{\mathrm{compute}}$ is a utilization factor reflecting kernel launch overheads, pipeline stalls, and non-ideal compute scheduling. This latency model helps determine if performance is limited by computational capacity or by data transfer bottlenecks. 
By adjusting $U_{\mathrm{compute}}$, users can model performance improvements from kernel fusion, operator scheduling, or optimized hardware utilization techniques such as Tensor Cores or systolic arrays in edge accelerators.

\paragraph{Memory latency} 
Memory latency computed as:
\begin{equation}
\label{mem}
  T_{\mathrm{mem}} =
  \frac{M}{\mathrm{memory}_{\mathrm{BW}} \times U_{\mathrm{memory}}}
\end{equation}

Equation \ref{mem} models the time to read or write total data volume $M$ across the device’s memory subsystem. $\mathrm{memory}_{\mathrm{BW}}$ is the theoretical bandwidth, and $U_{\mathrm{memory}}$ is utilization efficiency capturing factors such as bank conflicts, non-coalesced accesses, and DRAM refresh penalties. This estimation is essential for memory-bound workloads, where data movement rather than arithmetic dominates runtime. It guides decisions on operator reordering, memory tiling, quantization to reduce data size, and architectural choices such as increasing on-chip SRAM to reduce external DRAM usage.

\paragraph{I/O, Host-to-Device, and Network Overheads}
Estimates of data transfer overheads in critical edge deployments:
\begin{equation}
\label{I/O}
  T_{\mathrm{I/O}}   = \frac{P\,B}{\mathrm{STORAGE}_{\mathrm{BW}} \times U_{\mathrm{storage}}}
\end{equation}
\begin{equation}
\label{h2d}
  T_{\mathrm{h2d}}  = \frac{P\,B}{\mathrm{H2D}_{\mathrm{BW}} \times U_{\mathrm{H2D}}}
\end{equation}
\begin{equation}
\label{net}
  T_{\mathrm{net}}  = \frac{S\,H\,B}{\mathrm{NET}_{\mathrm{BW}} \times U_{\mathrm{net}}}
\end{equation}

Equations \ref{I/O}-\ref{net} estimate data transfer overheads. Equation \ref{I/O} models loading weights from storage (e.g. SSD or flash), \ref{h2d} models copying weights to device memory over PCIe or NVLink, and \ref{net} models exchanging key/value cache shards over a network in distributed settings. Each uses bandwidth and utilization terms to convert bytes to latency in milliseconds. These overheads are critical for edge systems where PCIe Gen3/Gen4 or embedded interconnect bandwidths can become bottlenecks.

\subsection{Energy Modeling}
EdgeProfiler models energy consumption
by \textit{Total energy per token} as:
\begin{equation}
\label{energy}
  E = \mathrm{FLOPs/token}\,\times\,e_{\mathrm{flop}}
    \;+\; M \,\times\,e_{\mathrm{byte}}
\end{equation}

Equation \ref{energy} computes token-level energy as the sum of compute energy (FLOPs $\times$ energy per FLOP) and memory energy (total bytes moved $\times$ energy per byte). $e_{\mathrm{flop}}$ and $e_{\mathrm{byte}}$ are hardware-specific constants obtained from power modeling or direct measurement. This metric is critical for battery-constrained edge deployment, enabling rapid estimation of average and peak energy consumption per inference. 


\begin{figure*}[t]
\centerline{\includegraphics[width=\textwidth]{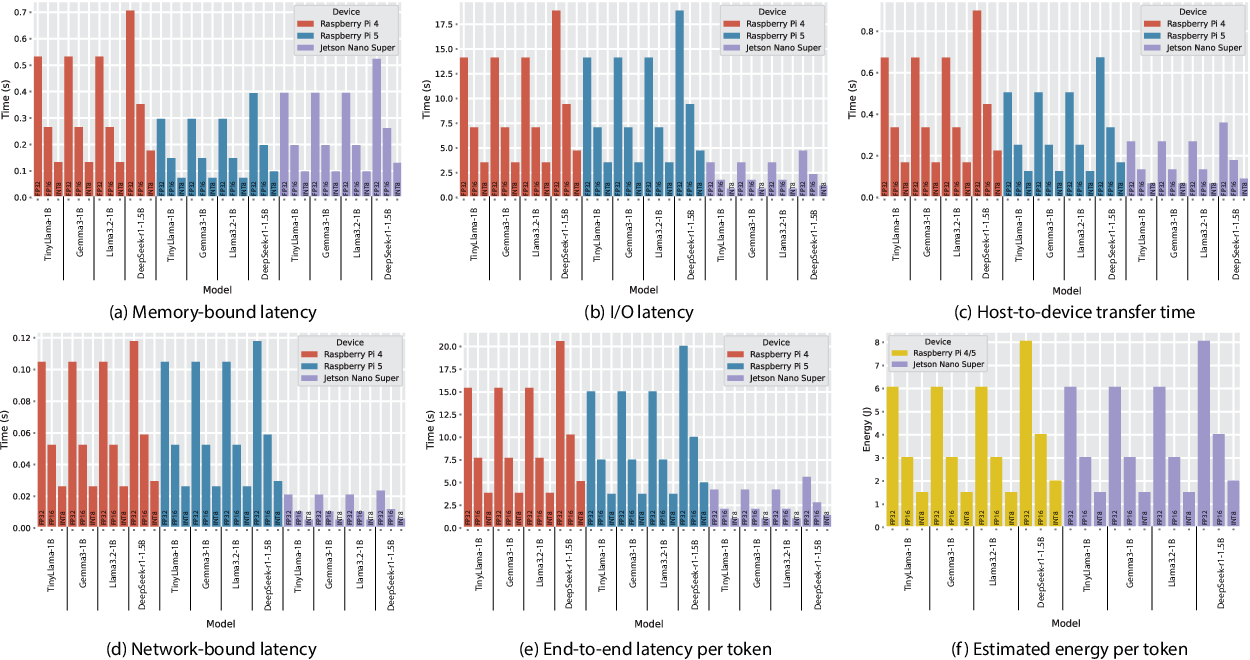}}
\caption{Latency and energy profile using EdgeProfiler: (a) Memory-bound latency, (b) Storage I/O latency to load weight from disk, (c) Host-to-device transfer time, (d) Network-bound latency for a single KV-shard exchange across nodes, (e) End-to-end latency per token, (f) Estimated energy per token.}
\label{letency_profile}
\end{figure*}

\begin{table*}[t]
\centering
\caption{Comparison of model size, memory usage, inference speed, and accuracy loss across different quantization precisions.}
\label{tab:quantization-comparison}
\begin{tabular}{lccccc}
\toprule
\textbf{Model} &\textbf{Precision} & \textbf{Model Size} & \textbf{Memory at Runtime} & \textbf{Inference Speed} & \textbf{Accuracy Loss} \\
\midrule

         &FP16 & 2.2GB & $\sim$3.13GB & 1$\times$ & baseline \\

TinyLlama&INT8 & 1.2GB & $\sim$2.25GB & 1.86$\times$ & minor \\

         &INT4 & 644MB & $\sim$1.78GB & 2.45$\times$ & moderate \\
\midrule

         &FP16 &2.0GB  & $\sim$2.44GB & 1$\times$ & baseline \\

Gemma3-1B&INT8 &1.1GB  & $\sim$1.60GB & 1.26$\times$ & minor \\

         &INT4 &815MB  & $\sim$1.35GB & 1.52$\times$ & moderate \\
\midrule

            &FP16 &2.5GB  & $\sim$3.58GB & 1$\times$ & baseline \\

Llama3.2-1B&INT8 &1.3GB  & $\sim$2.53GB & 2.7$\times$ & minor \\

            &INT4 &776MB  & $\sim$2.01GB & 3.33$\times$ & moderate \\
\midrule

                &FP16 &3.6GB  & $\sim$3.91GB & 1$\times$ & baseline \\

DeepSeek-r1-1.5B&INT8 &1.9GB  & $\sim$2.55GB & 2.19$\times$ & minor \\

                &INT4 &1.1GB  & $\sim$1.84GB & 2.97$\times$ & moderate \\

\bottomrule
\end{tabular}
\vspace{-10pt}

\end{table*}

\section{Evaluation}
\label{sec:eval}
This section evaluates the performance of lightweight LLMs on representative edge devices. The profiling framework is used to analyze trade-offs between model size, precision levels, and hardware characteristics. Key metrics such as latency, memory footprint, arithmetic intensity, and energy consumption are examined.
The framework was executed on a standard workstation equipped with a 10$^{th}$ Gen Intel(R) Core(TM) i7-10700F CPU, 32 GB DDR4 memory, and without using a dedicated GPU, running Ubuntu 22.04 LTS. EdgeProfiler is available on GitHub:\href{https://github.com/ShakyaJayakody/EdgeProfiler}{EdgeProfiler}

\medskip
\noindent
\textbf{Experimental Setup.}
We instantiate the profiler on three edge platforms: Raspberry Pi 4, Raspberry Pi 5, and Jetson Nano Super, using published peak FLOPs and bandwidths with calibrated utilization factors. The analysis covers multiple numeric precisions (FP32, FP16, INT8) and representative 1–1.5B parameter LLMs. For each configuration, the profiler outputs key metrics, including parameter count, FLOPs per token, peak memory footprint, stage-wise latency (compute, memory, I/O, H2D, network), end-to-end latency, arithmetic intensity, and energy per token. This analytical approach enables rapid comparison of architecture and precision trade-offs without requiring full hardware deployment. Details of the hardware configurations are summarized in Table~\ref{tab:device-comparison-os}.

\medskip
\noindent
\textbf{Profiling Results  Analysis.}
On all three devices, storage I/O accounts for the vast majority of end-to-end latency shown in Fig.~\ref{letency_profile}(b). Even though compute (and memory) times are on the order of a few hundred milliseconds or less, I/O delays range from multiple seconds (Raspberry Pi 4/5) down to just under a second on the Jetson Nano Super. This indicates that, without specialized weight-loading optimizations, simply reducing arithmetic cost (e.g., via quantization) will yield diminishing returns once compute time becomes negligible relative to data-movement overhead.

Precision reduction from FP32 to FP16 halves each component’s latency, and INT8 cuts it roughly by four. On Raspberry Pi 4, end-to-end latency drops from $\sim$15.4 s (FP32) to $\sim$3.9s (INT8), driven almost entirely by shorter I/O and transfer times of smaller weight footprints. However, even at INT8, I/O remains the bottleneck (3.5s vs. compute ~0.13s), underscoring that the network and storage subsystems must be improved in tandem with quantization to realize truly low-latency inference shown in all the Fig.~\ref{letency_profile}. The Jetson Nano Super’s higher storage bandwidth and PCIe host-to-device throughput, shown in Fig.~\ref{letency_profile}(c), deliver a dramatic reduction in I/O cost: INT8 inference completes in $\sim$1.05s end-to-end, nearly four times faster than on the Raspberry Pi 5. Compute and memory latencies on the Jetson ($\sim$0.07s and $\sim$0.88s at FP32) are comparable to—though still lower than those on the Pis, but the real advantage comes from overlapping and hiding I/O behind faster transfers. This highlights that mid-range AI accelerators can shift the bottleneck away from storage if paired with efficient weight-delivery mechanisms.


\medskip
\noindent
\textbf{Ablation Studies on Quantization.}
Table \ref{tab:quantization-comparison} summarizes that reducing precision from FP16 to INT8 and INT4 impacts model size, peak memory usage, inference throughput. We observe accuracy degradation across four 1-1.5B (TinyLlama‐1B, Gemma3‐1B, Llama3.2‐1B, DeepSeek‐r1‐1.5B) and peak runtime memory (3.13GB to 2.25GB) but a boost in inference speed by roughly $1.8\times$ with only a minor accuracy loss. Further quantization to INT4 yields even more storage saving (644MB for  TinyLlama) and a $2.45\times$ speedup, at the cost of a moderate drop in task performance. Across all architectures, INT8 delivers near-$2\times$ speed gain and $\sim$50\% reduction in memory footprint with negligible impact on accuracy. The large models Llama3.2-1B and DeepSeek-r1-1.5B benefit even more from low-precision storage, up to $3.3\times$ speedup, incur more noticeable accuracy degradation, suggesting their suitability only when memory and latency constraints are extreme and a moderate loss in quality is acceptable.

\section{Conclusion}
\label{sec:conclusion}

In this paper, we have propose EdgeProfiler, a fast profiling framework for lightweight LLMs on edge devices using an analytical model. 
EdgeProfiler framework reveals how hardware characteristics, model size, and numerical precision jointly shape inference performance and efficiency. We present our findings from three widely used edge devices.
Across all platforms, reducing precision from FP32, FP16, and INT8 yields substantial end-to-end latency and energy savings. 
On low-power devices, Raspberry Pi, memory, and I/O bound stage often dominate, whereas on more capable hardware like Jetson Nano, compute latency becomes negligible, and I/O or network transfer surface as the new bottleneck. Building upon this foundation, future work will focus on refining profiling granularity, integrating real-world deployment feedback, and expanding support for emerging hardware platforms to accelerate practical, scalable, and energy-efficient LLM applications on the edge.


\bibliographystyle{unsrt}
\bibliography{IEEEabrv,refs}

\end{document}